\documentstyle[twocolumn,aps]{revtex}
\def\Journal#1#2#3#4{{#1} {\bf #2}, #3 (#4)}
\input{epsfig.sty}
\def\NPA{{\rm Nucl. Phys.} A}
\def\NPB{{\rm Nucl. Phys.} B}
\def\PLB{{\rm Phys. Lett.}  B}
\def\PRL{\rm Phys. Rev. Lett.}
\def\PRD{{\rm Phys. Rev.} D}
\def\PRC{{\rm Phys. Rev.} C}

\def\la{\langle}
\def\ra{\rangle}
\def\be{\begin{equation}}
\def\ee{\end{equation}}
\def\bea{\begin{eqnarray}}
\def\eea{\end{eqnarray}}
\def\lsim{\mathrel{\rlap{\lower4pt\hbox{\hskip1pt$\sim$}}
    \raise1pt\hbox{$<$}}}         %less than or approx. symbol
\def\gsim{\mathrel{\rlap{\lower4pt\hbox{\hskip1pt$\sim$}}
    \raise1pt\hbox{$>$}}}
\begin{document}
\title{Skewed quark distribution of the pion in the light-front quark model}
\author{ Ho-Meoyng Choi$^{a}$, Chueng-Ryong Ji$^{b}$ 
and L.S. Kisslinger$^{a}$\\
 $^a$ Department of Physics, Carnegie-Mellon University\\
Pittsburgh, PA 15213\\
 $^b$Department of Physics, North Carolina State University\\
Raleigh, NC 27695-8202}
\maketitle
%\narrowtext
%\vspace{1.0in}
\begin{abstract}
We calculate the skewed quark distributions(SQDs) of the pion in the 
light-front quark model, and discuss the calculation of the nonvalence 
contribution to the SQDs in this model.  The frame-independence of our
model calculation is guaranteed by the constraint of the sum rule 
between the SQDs and form factor. 
Our numerical results show large nonvalence contributions
to the SQDs at small momentum transfer region as the skewedness 
increases.
\end{abstract}
PACS number(s): 13.40.Gp, 13.60.Fz, 12.39.Ki
\date{}
%\newpage
\section{Introduction}
Recently there has been a great interest in the 
off-forward (or nonforward, off-diagonal) parton 
distribution functions defined as non-diagonal hadronic matrix elements 
of bi-local products of the light-front quark and gluon field 
operators~\cite{Mu,XJ1,Ra1,Col,PW,Bl,Dh1,BDH,Bur1,DFJK}. 
The off-forward parton 
distribution functions, the so called ``skewed parton distributions (SPDs)",
are the generalization of the ordinary (forward) distribution 
functions. A well-known and practical example of SPDs as a nonperturbative 
information entering the light-front dominated hard scattering processes  
is the deeply virtual Compton scattering (DVCS) $\gamma^* p\to\gamma p$ for  
large initial photon virtuality $Q^2$ and small $t$ region, 
which can be factorized into a hard photon-parton and a skewed parton 
distribution~\cite{Mu,XJ1,Ra1}. 

In the present work we formulate the pion form factor in terms of the SPDs.
Since the usual local photon vertex in the pion form factor analysis is
replaced by a nonlocal operator of the SPDs, one can explore new 
physics. The physical interpretation of the SPDs becomes clear 
in the light-front frame with the light-front gauge $A^+=0$. 
In the light-front coordinates, the SPDs are in general functions
of the longitudinal momentum fraction variable $x$, the skewedness parameter
$\xi=(P-P')^+/P^+$ measuring asymmetry between initial ($P$) and 
final ($P'$) hadron state momenta, 
and the squared momentum transfer $t$. A simple physical 
interpretation of the SPDs, in light-front quantization, is that they provide 
a link between the ordinary parton distributions of hadrons and the hadronic
form factors, i.e. the ordinary parton distributions
are forward ($\xi=0$ and $t=0$) limits of the SPDs and the form 
factors are given by moments of them.

Due to this dual role of SPDs, they are closely related to  
form factors with the only difference between SPDs and form factors
being that the momentum of the ``probed quark" in SPDs is not integrated 
over but rather kept fixed at momentum fraction $x$.
For example, in studying the light-front wave functions of hadrons,
the overlap representation of the light-front wave functions of 
hadrons for the form factors in spacelike ($q^2<0$) region can be obtained
if one uses $J^+(=J^0+J^3$) and the Drell-Yan-West ($q^+=0$) frame where only 
parton-number-conserving valence Fock state contribution is needed.
The successful phenomenological calculations of form factors in spacelike 
region can be found in the light-front quark 
model(LFQM)~\cite{Ja,CJ1,CJ2,KCJ}.
On the other hand, form factors in timelike ($q^2 >0$) region 
such as the weak form factors for exclusive semileptonic decays
require $q^+>0$ frame, which in turn require parton-number-changing
nonvalence Fock state contribution as well as the valence 
one~\cite{BH,JC}. Similarly, while the ordinary parton 
distributions (analogous to $q^+=0$ limit of form factor) can 
be represented in terms of squared light-front wave function of a hadron, 
one cannot avoid nonvalence contributions to SPDs since 
they always involve non-zero $\xi$ corresponding to $q^+\neq 0$
in timelike form factor calculations.  
In recent papers~\cite{BDH,DFJK}, the nonvalence contribution to 
the SPDs has been rewritten in terms of light-front wave functions with 
different parton configurations. However, the representation given 
by~\cite{BDH,DFJK} requires to
find all the higher Fock-state wave functions while there has been
relatively little progress in computing the basic wave functions of
hadrons from first principles. 
Our approach provides an alternative way of handling the nonvalence
contribution which is more suitable for the constituent quark model(CQM)
specific to the low momentum transfer processes. We use
the light-front Bethe-Salpeter (B-S) formalism for the SPDs as an
extension of our treatment of the pion form factor~\cite{KCJ}.
Although the present work has many features similar to that of the
Fock state expansion, we utilize the close relation in the B-S formalism
between what is interpreted as the valence and nonvalence wave functions in
the Fock state approach.

In an effort to apply light-front
wave function based phenomenology to form factors in timelike exclusive
processes, we have presented in~\cite{JC} an effective treatment 
of handling the nonvalence contribution to the weak form factors, based
on the B-S formalism, and obtained reasonably good
numerical results for the processes in the small momentum transfer 
region. The main purpose of the present work is to apply the effective 
method presented in~\cite{JC} of handling the nonvalenece contribution 
to the SPDs of pion at small momentum transfer region in LFQM. 
The paper is organized as follows. In Section II, we briefly introduce
the necessary kinematics in which we follow the notation
employed by Radyushkin~\cite{Ra1}. In Section III, we represent the
SPDs of the pion in terms of light-front vertex functions, starting from
the covariant Bethe-Salpeter model of ($3+1$)-dimensional fermion field 
theory. The nonvalence part of SPDs is expressed in terms of 
light-front vertex functions of a hadron and a gauge boson.
The link operator connecting $(n-1)$-body to $(n+1)$-body in a Fock state
representation is obtained by an analytic continuation of the usual B-S 
amplitude. We also show that the complicated $(n+2)$-body energy denominators 
are absent through the calculation of the light-front time-ordered diagrams.
Of particular interest, the instantaneous contribution of the quark
propagator to the nonvalence diagram is seperated from the on-shell 
propagating part. In Section IV, we replace the light-front 
vertex functions obtained from Sec.~III with our LFQM wave 
function~\cite{CJ1} and show our numerical results for the SPDs 
of the pion at small momentum transfer region. 
We also show that the frame-independence of our model is guaranteed 
by the sum rule between the SPDs and form factor of the pion.
Conclusions follow in Section V. 
\section{Kinematics}

\begin{figure}
\centerline{\psfig{figure=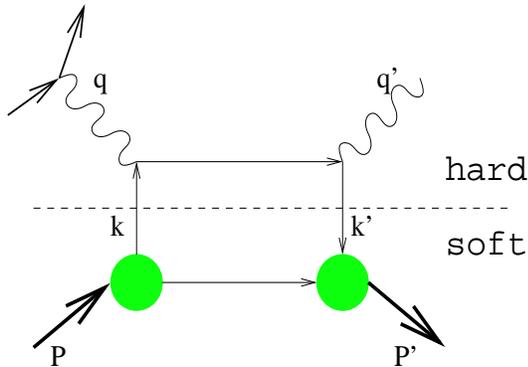,height=5cm,width=7cm}}
%\centerline{\psfig{figure=Fig1.eps,height=6cm,width=8cm}}
\caption{Handbag diagram contributing dominantly to Compton scattering in
the deeply virtual region. The lower soft part consists of a hadronic matrix
element which is parametrized in the form of skewed parton distribution 
functions.\label{handbag}}
\end{figure}

We begin with the kinematics of the virtual Compton 
scattering (see Fig.~\ref{handbag}) of the pion 
\be\label{VCS}
\gamma^*(q) + \pi(P) \to \gamma(q') + \pi(P'),
\ee
where the initial (final) hadron state is characterized by the momentum 
$P\;(P')$ and the incoming spacelike virtual and outgoing real photon 
momenta by $q$ and $q'$, respectively.  
We shall use the component notation 
$V=(V^+,V^-,{\bf V}_\perp)$ and our metric is specified by 
$V^{\pm}=(V^0 \pm V^3)$ and $V^2=V^+V^- - {\bf V}_{\perp}$.

Defining the four momentum transfer $\Delta=P-P'$, one has
\bea\label{PP}
P &=&\biggl[P^+, \frac{M^2}{P^+}, 0_\perp \biggr],\nonumber\\
P'&=&\biggl[(1-\xi)P^+,
\frac{M^2+\Delta^{2}_{\perp}}{(1-\xi)P^+}, -\Delta_\perp\biggr],
\eea
and
\be\label{del}
\Delta = P - P' =\biggl[
\xi P^+, \frac{\Delta^2+\Delta^2_\perp}{\xi P^+}, \Delta_\perp \biggr],
\ee
where $M$ is the pion mass and $\xi=\Delta^+/P^+$ is
the skewedness parameter describing the asymmetry in plus momentum.
The squared momentum transfer then reads
\be\label{del2}
t = \Delta^2 = 2 P\cdot\Delta=-\frac{\xi^2 M^2 + \Delta^2_\perp}{1-\xi}.
\ee
Since $\Delta^2_\perp\geq 0$, $t$ has a minimum value 
$-t_{\min}=\xi^2 M^2/(1-\xi)$ at given $\xi$.
As shown in Fig.~\ref{handbag}, the parton emitted by the pion 
has the momentum $k$, and the one absorbed has the momentum $k'$.

As in the case of spacelike form factors, we choose a frame
where the incident spacelike photon carries $q^+=0$:
\bea\label{qq}
q &=& \biggl[0, \frac{({\bf q}_\perp + \Delta_\perp)^2}{\xi P^+}
+ \frac{\xi M^2 +\Delta^2_\perp}{(1-\xi) P^+},
{\bf q}_\perp \biggr],\nonumber\\ 
q' &=& \biggl[\xi P^+, 
\frac{({\bf q}_\perp + \Delta_\perp)^2}{\xi P^+},
{\bf q}_\perp +\Delta_\perp\biggr]. 
\eea
In deeply virtual Compton scattering (DVCS) where $Q^2=-q^2$ is large 
compared to the mass $M$ and $-t$, one obtains
\be\label{Bj}
\frac{Q^2}{2P\cdot q}=\xi,
\ee
i.e. $\xi$ plays the role of the Bjorken variable in DVCS. For a fixed
value of $-t$, the allowed range of $\xi$ is given by
\be\label{range}
0\leq\xi\leq\frac{(-t)}{2M^2}\biggl(
\sqrt{1 + \frac{4M^2}{(-t)}}-1\biggr).
\ee 

\section{Skewed quark distribution of the pion}
Analogous to the pion electromagnetic (EM) form factor calculation
\be\label{EM}
J^+(0)\equiv\langle P'|{\bar\psi(0)}\gamma^+\psi(0)|P \rangle
=F_{\pi}(t)(P+P')^+,
\ee
we define the skewed quark distributions(SQDs) 
${\cal F}_{\pi}(\xi,x,t)$ of a pion by 
\bea\label{SPD}
{\cal J}^+ &\equiv&
\int\frac{dz^-}{4\pi}e^{i x P^+z^-/2}
\la P'|{\bar\psi}(0)\gamma^+\psi(z)|P\ra|_{z^+={\bf z}_{\perp}=0} 
\nonumber\\
&=&{\cal F}_{\pi}(\xi,x,t)(P+P')^+,
\eea     
where  $z=(z^+,z^-,{\bf z}_\perp)$ in a light-front representation. 
Note that the path-ordered exponential of the gauge field,
${\cal P}\exp[i\int z^{\mu}A_{\mu}]$, required by gauge invariance 
in Eq.~(\ref{SPD}) does not appear in the light-front gauge $A^+=0$. 
As one can see from Eqs.~(\ref{EM}) and~(\ref{SPD}), 
the ${\cal F}_{\pi}$ involves one less integration than
the form factor $F_{\pi}$ due to nonlocality of the current matrix 
element. The SQDs display characteristics of
the ordinary(forward) quark distribution in the limit of $\xi\to 0$
and $t\to 0$, on the other hand,  the first moment of the SQDs
is related to the form factor by the following sum 
rules~\cite{XJ1,Ra1}:
\be\label{sum}
\int^1_{0}dx\; {\cal F}_{\pi}(\xi, x, t) = F_\pi(t),
\ee
where ${\cal F}_{\pi}(\xi, x, t)=e_u{\cal F}^u_{\pi}(\xi, x, t)
-e_d{\cal F}^{\bar d}_{\pi}(\xi, x, t)$ and we assume 
isospin symmetry($m_u=m_{\bar d}$) so that  
${\cal F}^u_{\pi}(\xi, x, t)={\cal F}^{\bar d}_{\pi}(\xi, x, t)$.
Note that Eq.~(\ref{sum}) is independent of $\xi$, which provides
important constraints on any model calculation of the SQDs.
In general, the polynomiality conditions for the moments
of the SQDs~\cite{Jsong,Ra2} defined by
\be\label{poly}
\int^1_{0}dx\; x^{n-1} {\cal F}(\xi, x, t) = F_n(\xi,t),
\ee
require that the highest power of $\xi$ in the polynomial 
expression of $F_n(\xi,t)$ should not be larger than $n$.
These polynomiality conditions are fundamental properties of
the SQDs which follow from the Lorentz invariance.
In case of a spin-1/2 composite system~\cite{XJ1,BDH}, where 
the helicity non-flip($H$) and helicity flip($E$) SQDs are involved, 
the second($n=2$) moment of each SQD yields the $\xi$ dependent
form factors of the energy-momentum tensor, although the sum of the moments 
produces the $\xi$-independent form factors of the energy-momentum tensor.
When the sum rule($n=2$) is extrapolated to $-t=0$~\cite{XJ1},
it provides the information on the total
quark(i.e. quark orbital) contribution to the nucleon spin.
For the spin-0 composite system like the pion, 
the situation is quite different because
only one SQD ${\cal F}(\xi, x, t)$ exists.
We discuss our numerical results of $F_n(\xi,t) (n=1,2,3)$ for the pion
in the next section (Section IV).

\begin{figure}
\centerline{\psfig{figure=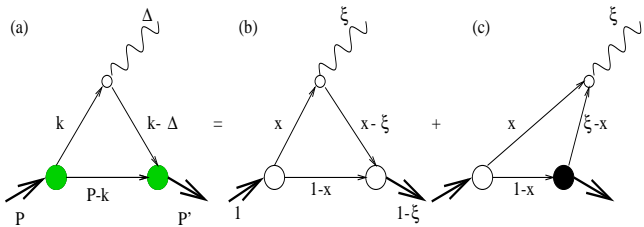,height=3cm,width=8.5cm}}
%\centerline{\psfig{figure=Fig2.ps,height=4cm,width=14cm}}
\caption{ Diagrams for SQDs in different kinematic regions for the
case $\xi>0$: The covariant diagram (a) corresponds to the sum of
the LF valence diagram (b) defined in $\xi< x<1$ region 
and the nonvalence diagram (c) defined in $0< x<\xi$ region.
The large white and black blobs at the meson-quark vertices in (b) and
(c) represent the ordinary LF wave function and the nonvalence 
wave function vertices, respectively.  
The small white blob at the quark-gauge boson 
vertex indicates the nonlocality of the vertex. 
\label{highFock}}
\end{figure}

In Refs.~\cite{BDH,DFJK}, the overlap representation of light-front
wave function for the SQDs has been obtained from the formal 
definitions of light-front field operators. In this work, however,
we shall derive the SQDs of the pion starting from
the covariant Bethe-Salpeter (B-S) amplitude of the current ${\cal J}$ 
given by (see Fig.~\ref{highFock}(a))
\bea\label{cu1}
{\cal J}^\mu &=& iN_c\int\frac{d^4 k}{(2\pi)^4}\frac{
\delta(x-k^+/P^+) H_{\rm cov}H'_{\rm cov} 
S^\mu}{[k^2-m^2+i\epsilon][(k-\Delta)^2-m^2+i\epsilon]}
\nonumber\\
&\times&\frac{1}{[(P-k)^2-m^2+i\epsilon]}, 
\eea
where $N_c$ is the color factor, $\delta(x-k^+/P^+)$ represents the
composite operator~\cite{Ra1} denoted by a small white blob 
at the quark-gauge boson vertex in Fig.~\ref{highFock},
and $H_{\rm cov}\;(H'_{\rm cov})$ is the covariant initial (final) state
meson-quark vertex function that satisfies the B-S equation. We refer to the
black vertex in Fig.~\ref{highFock}(c) as the nonvalence wave
function vertex as in the B-S formalism it is obtained by a continuation 
of the usual B-S amplitude, as we discuss below.

Using the following identity
\be\label{BL}
{\not\! p} + m = ({\not\! p}_{\rm on} + m)
+ \frac{1}{2}\gamma^+(p^- - p^{-}_{\rm on}),
\ee
we can express the trace term
$S^\mu$ in Eq.~(\ref{cu1}) in terms of the 
on-mass shell propagating part of quark propagators and the
instantaneous one as follows
\bea\label{trace}
S^\mu&=&{\rm Tr}\biggl[\gamma_{5}({\not\! p}_{1}+m)\gamma^\mu
({\not\! p}_{2}+m)\gamma_{5}(-{\not\!p}_{\bar{q}}+m)\biggr]
\nonumber\\
&=& {\rm Tr}\biggl[\gamma_{5}({\not\! p}_{\rm 1on}+m)\gamma^\mu
({\not\! p}_{\rm 2on}+m)\gamma_{5}(-{\not\!p}_{\bar{q}{\rm on}}
+m)\biggr]
\nonumber\\
&+& {\rm Tr\biggl[inst.\biggr]},
\eea
with
\bea\label{inst}
{\rm Tr\biggl[inst.\biggr]}&=&
2(p^{-}_{1} - p^{-}_{\rm 1on})
                  \biggl[p^{\mu}_{\rm 2on}p^{+}_{\bar{q}{\rm on}}
                  - p^{+}_{\rm 2on}p^{\mu}_{\bar{q}{\rm on}}
\nonumber\\
&&\hspace{2cm}
                  + g^{\mu+}(p_{\rm 2on}\cdot p_{\bar{q}{\rm on}}
                  + m^2) \biggr] \nonumber\\
&+& 2(p^{-}_{2} - p^{-}_{\rm 2on})
                  \biggl[p^{\mu}_{\rm 1on}p^{+}_{\bar{q}{\rm on}}
                  - p^{+}_{\rm 1on}p^{\mu}_{\bar{q}{\rm on}}
\nonumber\\
&&\hspace{2cm}
                  + g^{\mu+}(p_{\rm 1on}\cdot p_{\bar{q}{\rm on}}
                  + m^2) \biggr] \nonumber\\
&+& 2(p^{-}_{\bar{q}} - p^{-}_{\bar{q}{\rm on}})
                  \biggl[p^{\mu}_{\rm 1on}p^{+}_{\rm 2on}
                  + p^{+}_{\rm 1on}p^{\mu}_{\rm 2on}
\nonumber\\
&&\hspace{2cm}
                  - g^{\mu+}(p_{\rm 1on}\cdot p_{\rm 2on}
                  + m^2) \biggr] \nonumber\\
&+& 2g^{\mu+}p^{+}_{\bar{q}{\rm on}}
(p^{-}_{1} - p^{-}_{\rm 1on})(p^{-}_{2} - p^{-}_{\rm 2on}),
\eea
where $p_1=k,p_2=k-\Delta,$ and $p_{\bar q}=P-k$ and 
the subscript (on) means on-mass shell quark propagator.   

\begin{figure}
\centerline{\psfig{figure=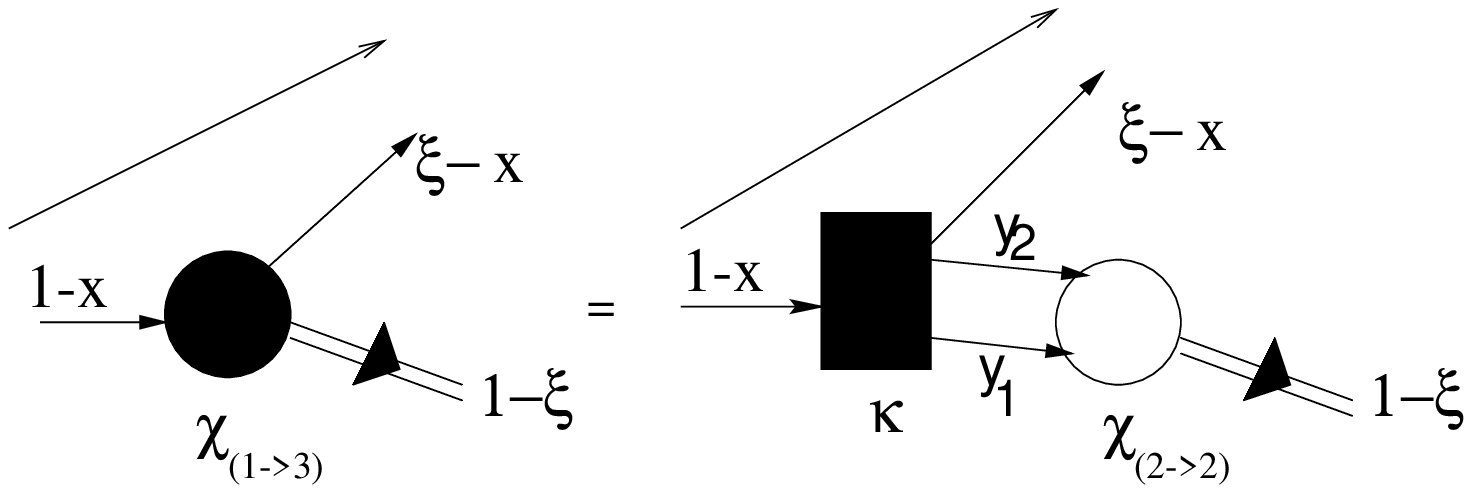,height=3cm,width=8cm}}
%\centerline{\psfig{figure=Fig3.ps,height=3cm,width=10cm}}
\caption{Nonvalence vertex(black blob) linked to an ordinary
light-front wave function(white blob).\label{SDFig}}
\end{figure}

The SQD is given by the solutions to the B-S 
equation~\cite{JC,BJS,Sales}
\bea\label{SDtype1}
&&(M^2_\xi-{\cal M}^2_0)\chi(x_i,{\bf k}_{i\perp})
\nonumber\\
&&=\int [dy][d^2{\bf l}_\perp]
{\cal K}(x_i,{\bf k}_{i\perp}; y_j,{\bf l}_{j\perp})
\chi(y_j,{\bf l}_{j\perp}),
\eea
where ${\cal K}$ is the B-S kernel which in principle includes all the 
higher Fock-state contributions, 
$M_\xi=M^2/(1-\xi)$, ${\cal M}^2_0=
(m^2+{\bf k}^2_{\perp})/(1-x) - (m^2 + {\bf k}^2_{\perp})/(\xi-x)$,
and $\chi(x_i,{\bf k}_{i\perp})$ is the B-S amplitude. Both the valence
and nonvalence B-S amplitudes are solutions to Eq.~(\ref{SDtype1}).
For the normal B-S amplitude, referred to as the valence wave function 
here, $x > \xi$, while for the nonvalence B-S amplitude $x < \xi$. We 
use the notation for these two solutions
\bea\label{BSamp}
   \chi_{(2\to 2)}& =& \chi^{val} \nonumber\\
 \chi_{(1\to 3)}& = & \chi^{nonval}.
\eea
This notation is motivated by the relationship to the Fock state 
picture, in which for the nonvalence vertex the parton number before and
after the kernel is interpreted as changing (from 1 to 3).
However, as illustrated in Fig.~\ref{highFock}(c),
the nonvalence B-S amplitude is an analytic continuation of the
valence B-S amplitude. In the LFQM the relationship between the
B-S amplitudes in the two regions is given by\cite{JC}
\bea\label{SDtype}
&&(M^2_\xi-{\cal M}^2_0)\chi_{(1\to 3)}(x_i,{\bf k}_{i\perp})
\nonumber\\
&&=\int [dy][d^2{\bf l}_\perp]
{\cal K}(x_i,{\bf k}_{i\perp}; y_j,{\bf l}_{j\perp})
\chi_{(2\to 2)}(y_j,{\bf l}_{j\perp}),
\eea  
where again the kernel includes in principle all the higher Fock-state
contributions because all the higher Fock components of the bound-state are
ultimately related to the lowest Fock component with the use of kernel.
This is illustrated in Fig.~\ref{SDFig}.

Equations (\ref{SDtype1}) and (\ref{SDtype}) are integral equations 
for which one needs nonperturbative QCD to obtain the kernel. 
We do not solve for
the B-S amplitudes in this work, but a nice feature of Eq.~(\ref{SDtype})
is a natural link between $\chi^{nonval}$ and $\chi^{val}$ which enables
an application of a light-front CQM even for the calculation of nonvalence
contribution in Fig.~\ref{highFock}(c).
We describe this below. In 1 + 1 QCD models~\cite{Bur1,Ein},
it is shown that expressions for the nonvalence vertex 
analogous to our form given in Eq.(\ref{SDtype}) are obtained.

  Corresponding to these two regions of the B-S amplitude,
the Cauchy integration over $k^-$ in Eq.~(\ref{cu1}) has two nonzero
contribution to the residue calculations, one coming from the interval
(I) $\Delta^+<k^+<P^+$ [see Fig.~\ref{highFock}(b)] and the other from
(II) $0<k^+<\Delta^+$ [see Fig.~\ref{highFock}(c)].

(I) In the region of $\Delta^+<k^+<P^+$, the residue is at the pole of 
$k^-=P^- - [{\bf k}^2_\perp + m^2-i\epsilon]/(P-k)^+$(i.e. 
$p^-_{\bar q}=p^{-}_{\bar{q}{\rm on}}$), which is placed in the
upper half of complex-$k^-$ plane. Thus, the Cauchy integration of 
${\cal J}^+$ in Eq.~(\ref{cu1}) over $k^-$ gives
\bea\label{jv}
{\cal F}^{val}_\pi(\xi,x,t)&=&
\frac{N_c}{(P+P')^+}\int^1_\xi
\frac{dx}{16\pi^3}\frac{\delta(x-k^+/P^+)}{x(1-x)x'}
\nonumber\\
&\times&
\int d^2{\bf k}_\perp
\chi_{(2\to 2)}(x,{\bf k}_\perp)S^+_{val}
\chi'_{(2\to 2)}(x',{\bf k}'_{\perp}),\nonumber\\
\eea 
where 
\bea\label{wfI}
\chi_{(2\to 2)}&=&\frac{h_{LF}}{M^2 -M^2_0},
M^2_0=\frac{{\bf k}^2_\perp + m^2}{1-x}
+ \frac{{\bf k}^2_\perp + m^2}{x},\nonumber\\
\chi'_{(2\to 2)}&=&\frac{h'_{LF}}{M^2 -M'^2_0},
M'^2_0=\frac{{\bf k'}^2_\perp + m^2}{1-x'}
+ \frac{{\bf k'}^2_\perp + m^2}{x'},\nonumber\\ 
\eea
and 
\be\label{SV}
S^+_{val}=\frac{4P^+}{(1-x')}({\bf k}_\perp\cdot{\bf k'}_\perp + m^2).
\ee
The internal momenta of the (struck) quark for the final state are 
given by
\be\label{xpkp}
x'= \frac{x-\xi}{1-\xi},\;\;
{\bf k'}_\perp={\bf k}_\perp + \frac{1-x}{1-\xi}\Delta_\perp.
\ee
While the light-front vertex function $h_{LF}\; (h'_{LF})$ formally 
is given by the covariant $H_{\rm cov}\; (H'_{\rm cov})$, in the 
present work the radial wave function 
$\chi_{(2\to 2)}\;( \chi'_{(2\to 2)})$
[consequently $h_{LF}\;(h'_{LF})$] is obtained from a light-front 
constituent quark model as we shall show later. 
It is also interesting to note that in this spectator pole diagram  
($p^-_{\bar q}=p^{-}_{\bar{q}{\rm on}}$) with the plus component of the
current, the instantaneous part in Eq.~(\ref{inst}) does not contribute
at all, i.e. all particles are on their mass shell. 

(II) In the region of $0<k^+<\Delta^+$, the residue is at the pole of
$k^-=[{\bf k}^2_\perp + m^2-i\epsilon]/k^+$ (i.e.
$p^-_1=p^{-}_{1{\rm on}}$), which is placed in the
lower half of complex-$k^-$ plane. Then the Cauchy integration of
${\cal J}^+$ in Eq.~(\ref{cu1}) over $k^-$ reads 
\bea\label{jnv}
&&{\cal F}^{nv}_\pi(\xi,x,t)\nonumber\\
&&\;=\frac{N_c}{(P+P')^+}\int^\xi_0
\frac{dx}{16\pi^3}\frac{\delta(x-k^+/P^+)}{x(1-x)x'}
\nonumber\\
&&\;\times\int d^2{\bf k}_\perp
\chi_{(2\to 2)}(x,{\bf k}_\perp)S^+_{nv}\chi^g(x,{\bf k''}_\perp) 
\nonumber\\
&&\;\times\int\frac{dy}{y(1-y)}\int d^2{\bf l}_\perp
{\cal K}(x,{\bf k}_\perp; y,{\bf l}_\perp)\chi_{(2\to 2)}(y,{\bf l}_\perp),
\nonumber\\
\eea
where Eq.~(\ref{SDtype}) has been used for the nonvalence wave function
at the black  blob in Fig.~\ref{highFock}(c),
and $\chi^g$ corresponds to the light-front energy denominator at the small
white blob in Fig.~\ref{highFock}(c). The explicit form of
$\chi^g$ is given by
\be\label{gaugeWF}
\chi^g(x,{\bf k''}_\perp)=
\frac{1}{(1-\xi)\biggl[\frac{\Delta^2}{\xi} - 
\biggl(\frac{{\bf k''}^2_\perp + m^2}{x} 
+\frac{{\bf k''}^2_\perp + m^2}{\xi - x}\biggr)
\biggr]}, 
\ee 
where ${\bf k''}_\perp = {\bf k}_\perp + (x/\xi)\Delta_\perp$.
We call $\chi^g$ the light-front vertex function of a 
gauge boson~\footnote{ While one can in principle also consider the B-S 
amplitude for $\chi^g$, we note that such extension does not alter our 
results within our approximation in this work because both hadron and gauge
boson should share the same kernel.}.
In the calculation of the trace term $S^+_{nv}$, one can easily see
from Eq.~(\ref{inst}) that there is one instantaneous contribution from 
the spectator ($p_{\bar q}$) line in addition to the on-mass shell 
contribution given by Eq.~(\ref{SV}), which leads to
\be\label{snv}
S^+_{nv}= S^{+}_{val} + \frac{4P^+}{1-x'}x(1-x)x'(M^2-M^2_0).
\ee

\begin{figure}
\centerline{\psfig{figure=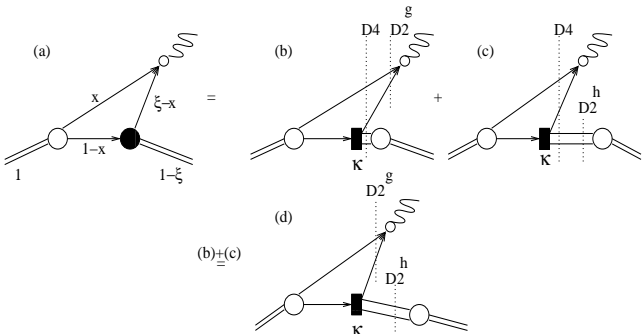,height=4.5cm,width=8.5cm}}
%\centerline{\psfig{figure=Fig4.eps,height=6cm,width=14cm}}
\caption{Effective treatment of the light-front nonvalence amplitude given
by Fig.~\protect\ref{highFock}(c). \label{Embed}}
\end{figure}

Considering the quark-meson and quark-gauge boson vertices together,
we also find that the four-body energy denominator ($D_4$)
appearing in Fig.~\ref{SDFig} is absent in Eq.~(\ref{jnv}). This
absence of $D_4$ in our nonvalence current matrix amplitude
given by Eq.~(\ref{jnv}) is due to the sum of two possible diagrams
in the light-front time-ordering (see Figs.~\ref{Embed}(b) and~(c)).
%The physical interpretation of those two diagrams are as follows:
%the $q{\bar q}$ pair annihilates into the current 
%after(see Fig.~\ref{Embed}(b)  or before(see Fig.~\ref{Embed}(c))
%merging into the final meson state. 
Summing over the two time-ordered 
diagrams Figs.~\ref{Embed}(b) and~(c), one can easily find the following 
identity, $1/D_4D^g_2+1/D_4D^h_2=1/D^g_2D^h_2$, which removes the complicate
four-body energy denominator term. We thus obtain the amplitude corresponding
to the nonvalence contribution given by Eq.~(\ref{jnv}) in terms of
ordinary light-front wave functions of hadron ($\chi_{(2\to 2)}$) and 
gauge boson ($\chi^g$) as shown in Fig.~\ref{Embed}(d). 
This method, however, requires the relevant
operator ${\cal K}(x,{\bf k}_\perp; y,{\bf l}_\perp)$ which is in general
dependent on the involved momenta connecting
the one-body to three-body sector as depicted in Fig.~\ref{SDFig}.
While the relevant operator ${\cal K}$
is in general dependent on all internal momenta 
$(x,{\bf k}_\perp; y,{\bf l}_\perp)$, the integral of ${\cal K}$
over $y$ and ${\bf l}_\perp$ in Eq.~(\ref{jnv}), which we define as
$G_\pi\equiv\int[dy][d^2{\bf l}_\perp]
{\cal K}(x,{\bf k}_\perp;y,{\bf l}_\perp)\chi_{(2\to 2)}(y,{\bf l}_\perp)$,
depends only on $x$ and ${\bf k}_\perp$. 
In this work, we approximate $G_\pi$ as a constant which
has been tested in our previous exclusive semileptonic decay 
processes~\cite{JC}
and proved to be a good approxiamtion at least for small momentum transfer 
region. As we shall show in the next section, the validity of this 
approximation can be checked by examining the frame-independence of 
our numerical results.

\section{Model Calculation of SQD}
In the previous sections we have derived the skewed quark distribution 
function ${\cal F}_\pi(\xi,x,t)$ starting from a covariant model. 
In this section we replace the light-front vertex function 
$h_{LF}$ (or equivalently $\chi_{(2\to 2)}(x,{\bf k}_\perp)$) by the standard
light-front vertex function~\cite{Ja,CJ1,CJ2}, which is symmetric in the 
variables of the constituent $q\bar{q}$ pair and has been successful in 
predicting many static properties of ground state mesons. 
Different choices of the vertex function are of course possible.

Comparing $\chi_{(2\to 2)}$ with our light-front
wave function given by Ref.~\cite{CJ1}, we identify 
\be\label{LFvertex}
\chi_{(2\to 2)}(x,{\bf k}_\perp)=
\sqrt{\frac{8\pi^3}{N_c}}
\sqrt{\frac{\partial k_z}{\partial x}}
\frac{[x(1-x)]^{1/2}}{M_0}\phi(x,{\bf k}_\perp),
\ee
where the Jacobian of the variable tranformation 
${\bf k}=(k_z,{\bf k}_\perp)\to
(x,{\bf k}_\perp)$ is obtained as $\partial k_z/\partial x=M_0/[4x(1-x)]$
and the radial wave function is given by 
\be\label{radial}
\phi({\bf k}^2)=
\sqrt{\frac{1}{\pi^{3/2}\beta^3}}
\exp(-{\bf k}^2/2\beta^2),
\ee
which is normalized as $\int d^3k|\phi({\bf k}^2)|^2=1$.   

Substituting Eqs.~(\ref{LFvertex}) and~(\ref{radial}) into 
Eqs.~(\ref{jv}) and~(\ref{jnv}), we obtain the valence and nonvalence
contributions to the SQDs of the pion in LFQM
\bea\label{light-frontFv}
{\cal F}^{val}_\pi(\xi,x,t)&=&
\frac{\theta(x-\xi)}{1-\frac{\xi}{2}}
\int d^2{\bf k}_\perp
\sqrt{\frac{\partial k'_z}{\partial x'}}
\sqrt{\frac{\partial k_z}{\partial x}}
\phi(x',{\bf k'}_\perp)\nonumber\\
&&\;\;\times
\phi(x,{\bf k}_\perp)\frac{({\bf k}_\perp\cdot{\bf k'}_\perp + m^2)}
{\sqrt{{\bf k}^2_\perp + m^2}\sqrt{{\bf k'}^2_\perp + m^2}}, 
\eea
and
\bea\label{light-frontFnv}
{\cal F}^{nv}_\pi(\xi,x,t)&=&\frac{\theta(\xi-x)}{1-\frac{\xi}{2}}
\int d^2{\bf k}_\perp
\sqrt{\frac{\partial k_z}{\partial x}}
\phi(x,{\bf k}_\perp)\chi^g(x,{\bf k''}_\perp)\nonumber\\
&\times& 
\frac{{\bf k}_\perp\cdot{\bf k'}_\perp + m^2 
+ x(1-x)x'(M^2-M^2_0)}
{x'(1-x')\sqrt{{\bf k}^2_\perp + m^2} }\nonumber\\
&\times&\int^1_0 dy\int d^2{\bf l}_\perp
\sqrt{\frac{{\partial l_z}}{{\partial y}}}
\frac{{\cal K}(x,{\bf k}_\perp; y,{\bf l}_\perp)}
{\sqrt{{\bf l}^2_\perp + m^2}}\phi(y,{\bf l}_\perp),\nonumber\\
\eea
where 
we treat the last term in Eq.~(\ref{light-frontFnv}) as a
constant $G_\pi$ which will be fixed by the sum rule expressed
in terms of ${\cal F}^{val}_\pi$ and ${\cal F}^{nv}_\pi$ as 
\be\label{2sum}
F_\pi(t)=\int^1_\xi dx\; {\cal F}^{val}_\pi(\xi, x,t)
+ \int^\xi_0 dx\; {\cal F}^{nv}_\pi(\xi, x,t),
\ee
for given $-t$. We note that Eq.~(\ref{2sum}) is used as a constraint
on the frame-independence of our model. 

In our numerical calculations, we use the model parameters 
$(m,\beta)=(0.22,0.3659)$ [GeV] obtained in Ref.~\cite{CJ1} for the 
linear confining potential model.
Before we calculate the SQDs and the form factor
of the pion, we first consider the valence contribution to the pion 
EM form factor and see how much the nonvalence contribution is
needed to obtain the frame-independent result of our model. 
To this end, we show in Fig.~\ref{piform} the valence contribution
(see Fig.~\ref{highFock}(b)) to the pion EM form factor with different values
of $\xi$, where the thick solid, cross (x), dotted, dot-dashed, and thin solid
lines represent the purely transverse ($\xi=0$ and $\Delta_\perp\neq 0$),
$\xi=0.02$, 0.3, 0.6, and the purely longitudinal (i.e. $\xi\neq 0$ and 
$\Delta_\perp=0$) results, respectively, and compare with the experimental
data~\cite{Amen,JLab}. Note that the purely transverse 
frame result (thick solid line) is the exact solution within our
model calculation. As one can see from Fig.~\ref{piform}, the nonvalence 
contribution to the pion form factor, i.e.
the difference between $\xi=0$ and $\xi\neq 0$ results,
increases as $\xi$ does. 
Our special interesting region in this work is the small $-t$ region where 
the nonvalence contribution is especially large and our effective method 
for the calculation of the nonvalence contributions works pretty well.    

Including the nonvalence contribution given by Eq.~(\ref{light-frontFnv})
to the SQDs of the pion, we can determine our constant 
$G_\pi$ in a frame-independent way by the sum rule
$\int^1_0 dx\;{\cal F}^{val}_\pi(\xi=0,x,t)=
\int^1_\xi dx\;{\cal F}^{val}_\pi(\xi\neq 0,x,t)
+ \int^\xi_0 dx\;{\cal F}^{nv}_\pi(\xi\neq 0,x,t)=F_\pi(t)$.
In Fig.~\ref{Gfactor}, we show the $\xi$-dependence of 
$G_\pi$ for different $-t$-values, i.e. $-t=0$ (diamond), 0.2 (black
circle), 0.5 (white circle), and  1.0 (black square) [GeV$^2$], respectively. 
As one can see in Fig.~\ref{Gfactor}, $G_\pi$ shows approximately 
constant behavior for $\xi>0.1$ at given small $-t$.
It is not surprising to see that $G_\pi$ becomes very large
as $\xi\to 0$, however, this does not cause a significant error in our
calculation because the nonvalence contribution in the very small $\xi$
region is highly suppressed. 
Note from Eq.~(\ref{jnv}) that ${\cal F}^{nv}_\pi$ has
the form of ${\cal F}^{nv}_\pi = G_\pi \times \int^\xi_0...$; thus
$G_\pi$ must be very large as $\xi\to 0$ to give the small contribution
of ${\cal F}^{nv}_\pi$, since the integral vanishes. Therefore the
results are consistent with an almost constant value for $G_\pi$
at least for small $-t$ as we show below. 
On the other hand, we note that there is an obvious $t$-dependence for
$G_\pi$, which might be the limit of our constant approximation.
In principle, we can obtain the SQDs in a frame-independent
way by using the true values of $G_\pi$ as shown in Fig.~\ref{Gfactor}
for given $(\xi,t)$. 
In the following, we compare the SQDs and the form factor
obtained from true values of $G_\pi$ (i.e.
frame-independent result of our model) with those
obtained from a single average value of $G_\pi= G_{\rm ave.}=0.32$ for
all $(\xi,t)$ to check the reliability of our constant $G_\pi$ approximation. 

In Figs.~\ref{SQD02} and~\ref{SQD1}, we show the SQDs 
${\cal F}_\pi(\xi,x,t)$ of the pion for fixed momentum transfer
$-t=0.2$ GeV$^2$ ($0\leq\xi\leq 0.92$) and
$-t=1.0$ GeV$^2$ ($0\leq\xi\leq 0.98$)
but with different skewedness parameters $\xi$, respectively.   
The solid and cross (x) lines in the nonvalence contributions 
are the exact solutions obtained from true values of $G_\pi$ and
the effective ones obtained from our average value of $G_{\rm ave.}=0.32$,
respectively. The dotted lines represent the instantaneous
contributions to ${\cal F}^{nv}_\pi(\xi,x,t)$ obtained from true
values of $G_\pi$. The SQDs at $\xi=0$ as shown in
Figs.~\ref{SQD02}(a) and~\ref{SQD1}(a) correspond to the ordinary 
quark distributions with vanishing nonvalence contributions.
The frame-independence
of our model calculation is ensured by the
area under the solid lines(valence $+$ nonvalence) being equal to the
pion form factor at given $-t$.
As one can see from Figs.~\ref{SQD02}(b-c) and~\ref{SQD1}(b-c), while
the nonvalence contributions are small for small $\xi=0.3$, they
are large for large skewdness parameter $\xi=0.9$. 
In our model calculations, the nonvalence 
contributions obtained from true values of $G_\pi$ (solid lines in
each figure) at $-t$=0.2 (1.0) GeV$^2$ for $\xi$=0.3 (0.3) and 
0.9 (0.9) are approximately 11 (4) $\%$ and 90 (85) $\%$, 
respectively. Comparing with the exact solutions, the numerical results with 
a single average $G_{\rm ave.}=0.32$ (cross lines in each figure) are shown
to reproduce the exact ones up to  97 $\%$ for $\xi=0.3$ and 90 $\%$ for 
$\xi=0.9$, respectively. It is also interesting to note that the instantaneous
contribuitons (dotted lines in each figure) become more pronounced as 
$\xi\to\xi_{\rm max}$ for each $-t$, which is a very different feature 
from a scalar theory model~\cite{Bur1} where there is no such instantaneous 
contribution.
While the instantaneous part of the nonvalence contribution vanishes
as $x \rightarrow \xi^-=lim_{\epsilon\rightarrow 0}(\xi-\epsilon)$
as shown in Figs.~\ref{SQD02} and~\ref{SQD1}, the net result
of ${\cal F}^{nv}_\pi(\xi,x,t)$ including the on-mass shell propagating
part does not\footnote{In fact, this behavior of ${\cal F}^{nv}_\pi(\xi, 
x, t)$ at $x=\xi$ has been anticipated in~\cite{Bur1} without a proof.}
vanish as $x=\xi$ and consequently causes a discontinuity to the zero
value of ${\cal F}^{val}_\pi(\xi,\xi,t)$. 
However, such discontinuity at $x=\xi$ is just an artifact due to 
the difference in the $x\rightarrow\xi$ behavior between the gauge boson 
vertex ($\chi^g(x,{\bf k}''_\perp)$) in Eq.~(\ref{gaugeWF}) and the hadronic
vertex ($\chi'_{2 \to 2}(x',{\bf k}'_\perp)$ in Eq.~(\ref{jv}) of 
our approximate model calculation.
We have indeed confirmed that the discontinuity at $x=\xi$ does not
occur in the limit of a point hadron vertex as already noticed
in the QED calculation~\cite{BDH}\footnote{We are grateful to M. Diehl for 
discussion of this point.}.
Thus, for a full analysis of DVCS satisfying the factorization 
theorems\cite{Pet}, it would be necessary to solve the bound-state B-S 
equation similar to Eq.~(\ref{SDtype1}) for the gauge boson ($\chi^g$) as
well as for the hadron ($\chi_{2\to 2}$).
In the chiral quark-soliton model analysis of the nucleon SQDs~\cite{Pet},
the discontinuity at $x=\xi$ was imputed to an artifact of neglecting 
the momentum dependence of the constituent quark mass.
Both the B-S amplitude for $\chi^g$ and the dynamical quark mass would
be anyway necessary for the model improvement.
Nevertheless, our effective method seems useful for the present
study of the relation between SQDs and the form factor in the 
nonperturbative regions.
 
In Fig.~\ref{Piave}, we show our effective solutions of the pion
form factor for $\xi=0.3$ (thin solid line) and 0.9 (long-dashed line) 
cases obtained from our average value of $G_\pi=G_{\rm ave.}=0.32$ 
and compare with the exact solution with $\xi=0$ (thick solid line) as well
as the experimental data~\cite{Amen,JLab}. The dotted and dot-dashed lines
represent the valence  and the instantaneous contributions to the form
factor for the case of $\xi=0.9$, respectively. In fact, there are 
$-t_{\rm min}$ values for nonzero $\xi$ due to 
$\Delta^2_\perp\geq 0$ (see Eq.~(\ref{del2})). 
We thus use the analytic continuation by
changing $\Delta_\perp$ to $i\Delta_\perp$ in 
Eqs.~(\ref{light-frontFv}) and~(\ref{light-frontFnv}) to
obtain the result for $0\leq -t\leq -t_{\rm min}$ where there is
no singularity.
A continuous behavior of the form factor near $-t_{\rm min}$ confirms 
the analyticity of our model calculation.
Our effective solution (thin solid line) with $\xi=0.9$ shows 
almost maximum deviation ($\lsim$ 10 $\%$) from the exact one 
(thick solid line) and
the deviation becomes smaller as $\xi$ reduces.
Our effective method of evaluating the nonvalence diagram
with a constant operator $G_\pi$ shows a definite improvement to restore the
frame-independence of our model and seems to be a quite reliable 
approximation.

In Fig.~\ref{Mellin}, we show the $n$th moments($n=1,2,3$) of 
${\cal F}_\pi(\xi,x,t)$ given by Eq.~(\ref{poly}) at 
$t=0$ using the true value of $G_\pi$ shown in 
Fig.~\ref{Gfactor}. Although at $t$ = 0 the skewedness parameter vanishes
(see Eq.~(\ref{range})), we use the model forms for ${\cal F}^{val}_\pi$ 
and  ${\cal F}^{nv}_\pi$ given by 
Eqs.~(\ref{light-frontFv},\ref{light-frontFnv}) to define the extrapolation
to $\xi \neq 0$. The thick solid, dotted, and dot-dashed lines
represent the total(=valence + nonvalence), valence, and nonvalence
contributions, respectively. For comparison, we
also show the nonvalence contribution obtained from the average value
of $G_\pi=0.32$(cross lines). As one can see, the first($n=1$) moment
(top thick solid line) is $\xi$-independent because
the sum rule for $n=1$ yields the physical pion form 
factor, $F_1(\xi,t)=F_\pi(t)$.
Also, the higher moments $F_2(\xi,t)$(middle thick solid line)
and $F_3(\xi,t)$(bottom thick solid line) satisfy the polynomiality
conditions(See Eq.~(\ref{poly})) discussed in the previous section.
In  Fig.~\ref{Mellin} we also plot the phenomenological form
\bea\label{2nd}
&&F_n(\xi,t=0)\nonumber\\
&&\simeq F_n(\xi=1)-(1-\xi)^2[F_n(\xi=1)-F_n(\xi=0)],
\eea 
shown by the diamonds.
The numerical values for $F_n^v = F_n(\xi=0)$ and $F_n^{nv}=F_n(\xi=1)$
are summarized in Table~I. Our value of $F_3^v$ is in good agreement
with the value obtained by the QCD sum rules with non-local 
condensates~\cite{Bel}.

We also compute the $n$th moment of the $\pi$ wave function,
 $\phi_{\pi}(y) = {\cal F}_\pi(\xi=0,y,t=0)$,
defined by~\cite{CZ} 
\bea\label{ymom}
\la y_n\ra &=& \int^1_{-1}dy y^{n} \phi_{\pi}(y)
\eea
where $y=2x-1$.
Our numerical results are summarized in Table~II and compared with 
several other theoretical results. It is interesting to note that
our value of $\la y_2 \ra$ is very close to the asymptotic value ($0.2$)
of the second moment obtained from the well-known asymptotic quark
distribution amplitude $\phi_\pi (y) = \frac{3}{4}(1-y^2)$, which is
quite different from the early work obtained with QCD sum rules~\cite{CZ}
shown in the second row of Table~II. A later calculation using QCD sum 
rules~\cite{MR}, shown in the third row, found moments closer to the
asymptotic values.  However,
we emphasize that the Bethe-Salpeter amplitude used in the present work
is a model rather than a solution to the B-S equation with a kernel
derived from nonperturbative QCD.

Finally, in Fig.\ref{isoscalar}, we obtain the isosinglet
SQDs of the pion by subtracting the valence part ${\cal F}_\pi (\xi=0,x,t=0)$
from the nonvalence part ${\cal F}_\pi (\xi=1,x,t=0)$.
Our result (dotted line) in Fig.~\ref{isoscalar} is qualitatively
very similar to the total isoscalar skewed quark distribution in the
pion satisfying the soft pion theorem (Fig.~5 of Ref.~\cite{PW}),
that is obtained by the low-energy effective field theory based on
the instanton model of the QCD vacuum.

\section{Conclusion}
\begin{figure}[t]
\centerline{\psfig{figure=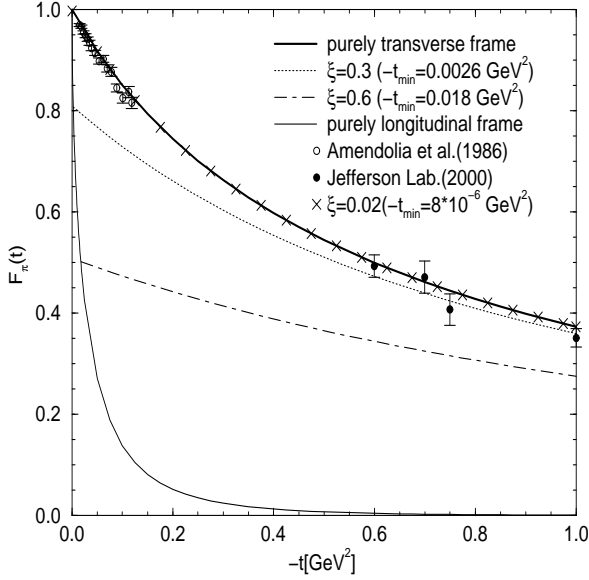,height=9cm,width=9cm}}
\caption{The valence contribution to the pion EM form factor with different
skewedness parameters $\xi$ compared with the experimental
data~\protect\cite{Amen,JLab}.\label{piform}}
\end{figure}

In this work, we investigated the SQDs of the pion
for small momentum transfer ($-t\leq 1$ GeV$^2$) region in the 
light-front quark model. 
Since the light-front nonvalence contributions to the SQDs 
of the pion are large 
especially at small momentum transfer region as shown in Fig.~\ref{piform}, 
it is very crucial to take them into account to guarantee the 
frame-independence of the model. 
Applying our effective treatment~\cite{JC},
i.e. the nonvalence B-S amplitude given by Eq.~(\ref{SDtype})
of the nonvalence contribution to the SQDs 
${\cal F}_\pi$ (=${\cal F}^{val}_\pi + {\cal F}^{nv}_\pi$) 
of the pion, we express ${\cal F}^{nv}_\pi$ (see Eqs.~(\ref{jnv}) 
and~(\ref{light-frontFnv})) in terms of ordinary 
light-front wave functions of a gauge boson and a hadron and  
calculate this nonvalence contribution numerically. 

\begin{figure}
\centerline{\psfig{figure=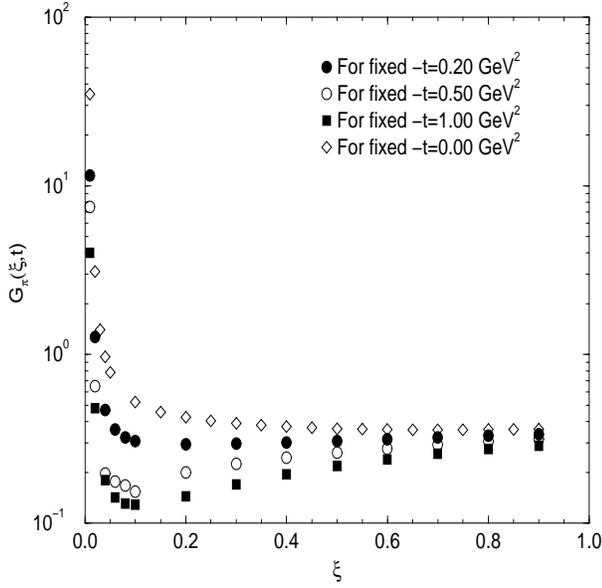,height=9cm,width=9cm}}
\caption{The $\xi$-dependence of $G_\pi$ for different
momentum transfers $-t=0$ (diamond), 0.2 (black circle),
0.5(white circle), and 1 (black square) [GeV$^2$], 
respectively. \label{Gfactor}}
\end{figure}

\begin{figure}[t]
\centerline{\psfig{figure=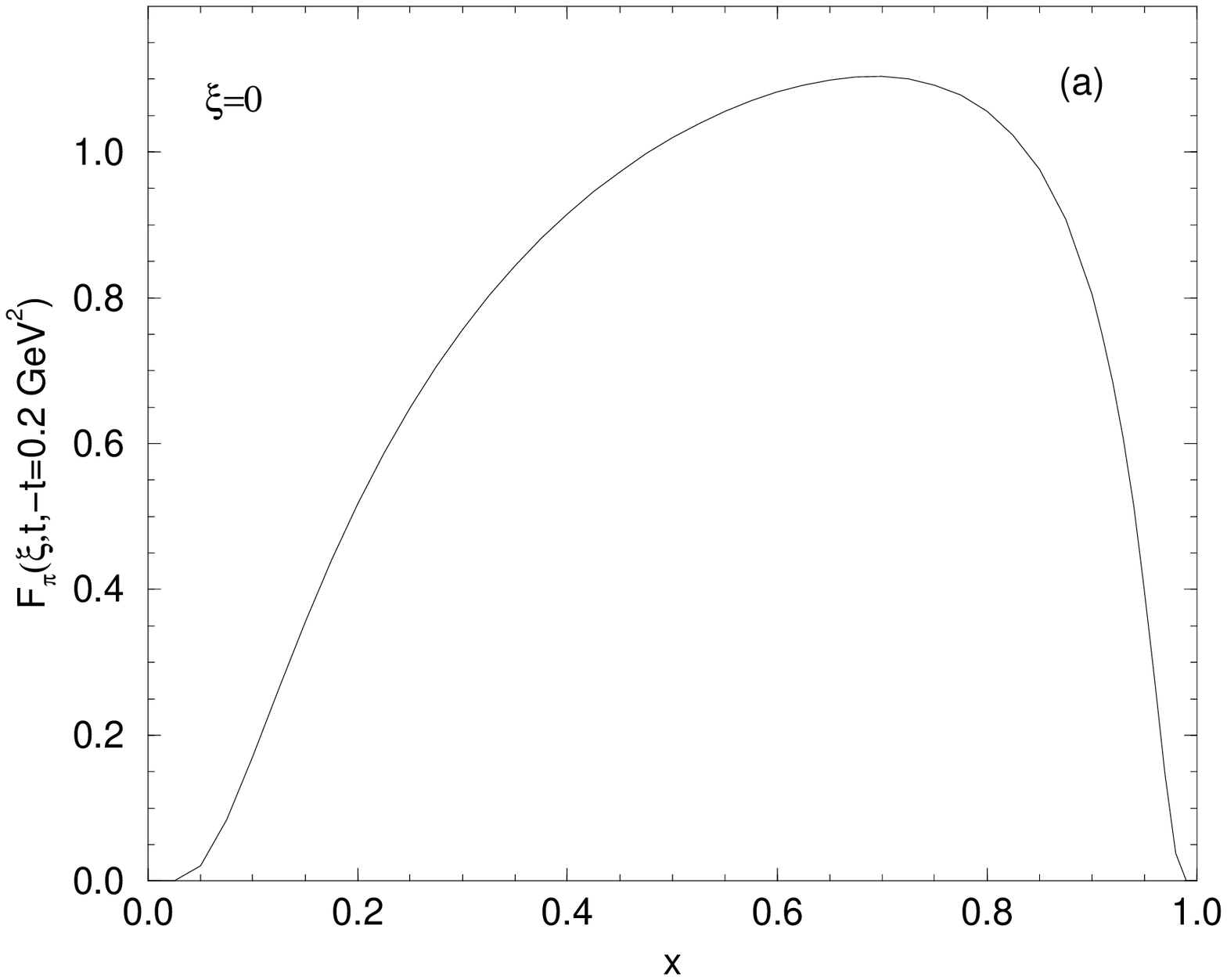,height=6.5cm,width=9cm}}
\centerline{\psfig{figure=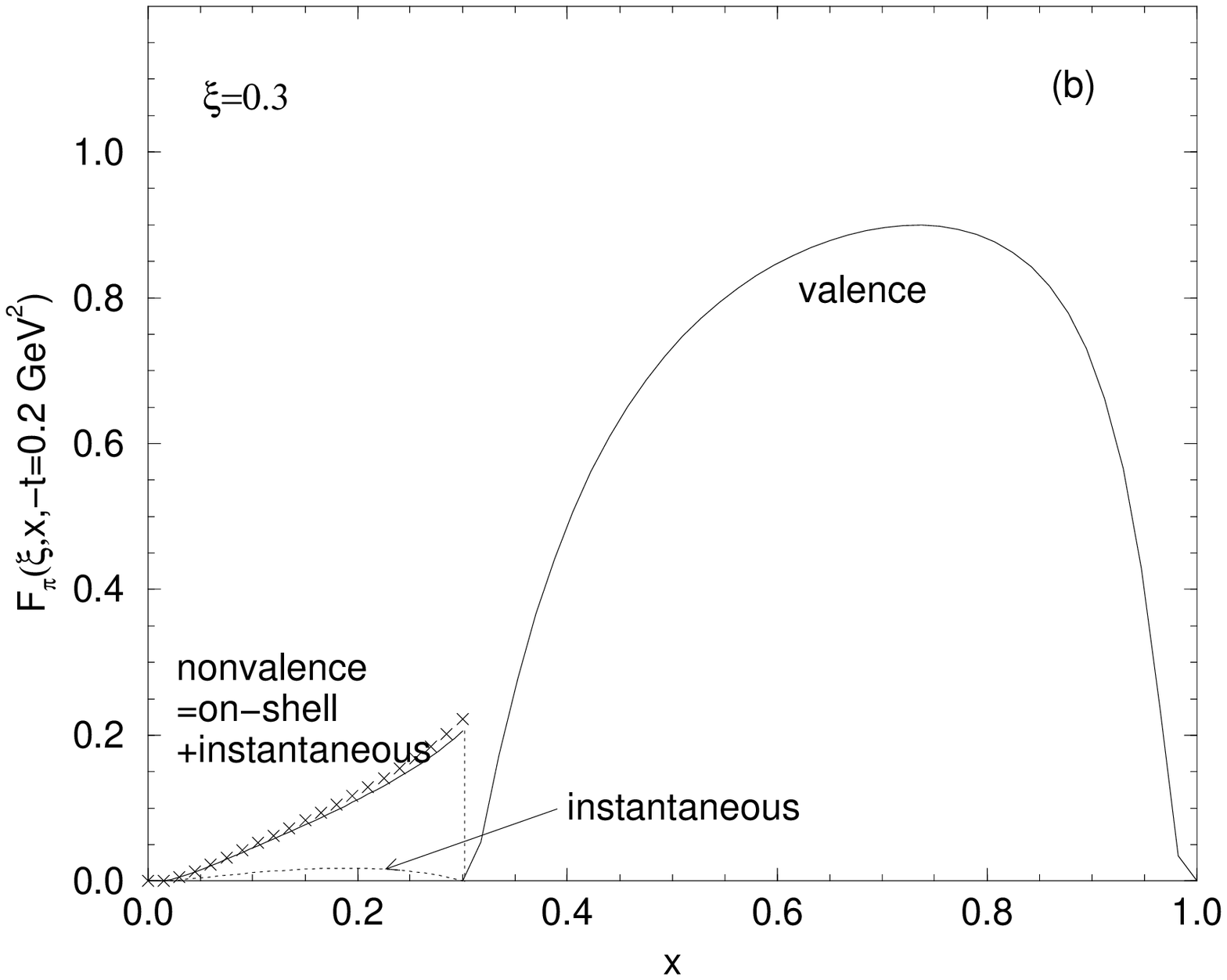,height=6.5cm,width=9cm}}
\centerline{\psfig{figure=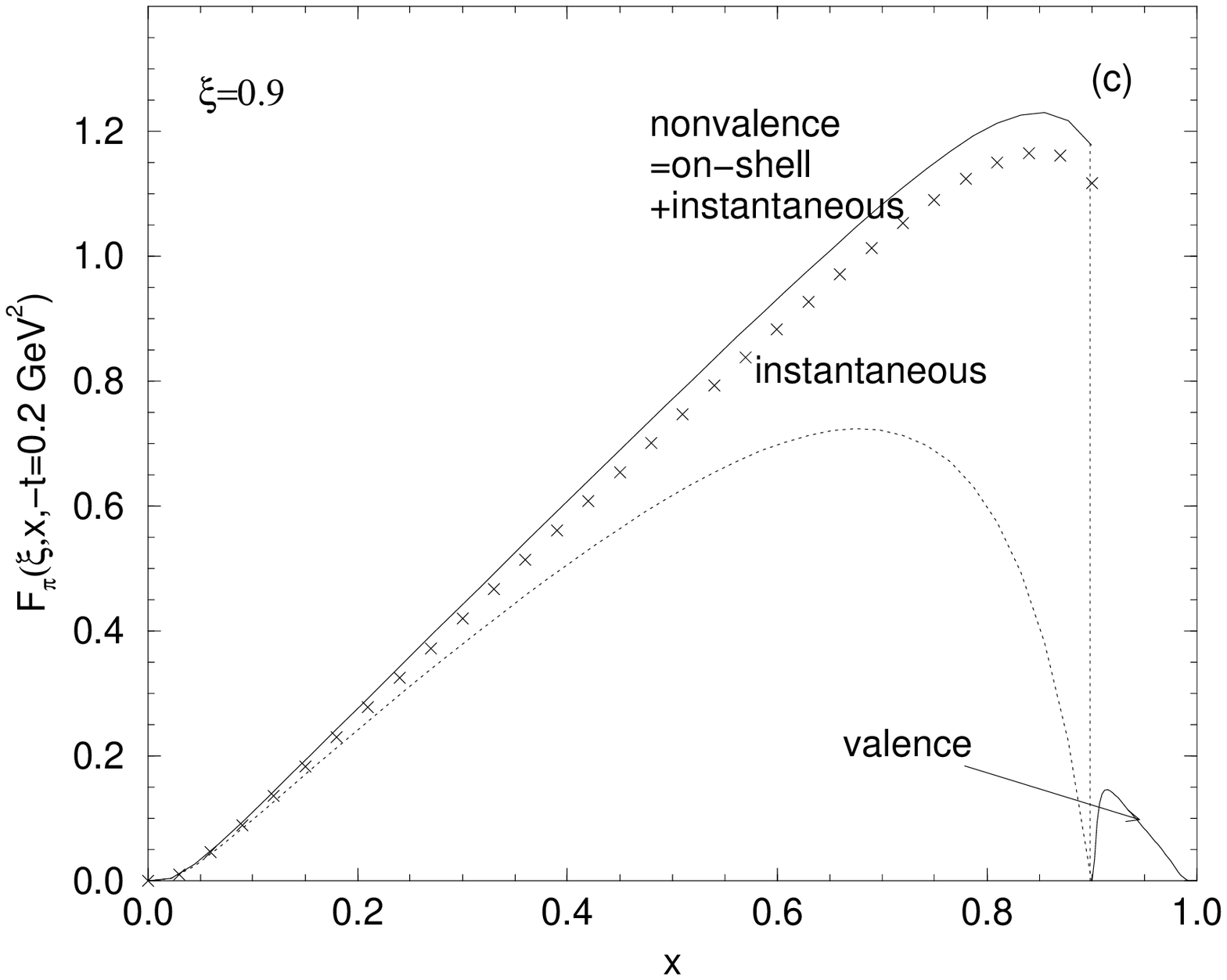,height=6.5cm,width=9cm}}
%\vspace{0.5cm}
\caption{Skewed quark distributions of the pion at
$-t=0.2$ GeV$^2$ with $\xi=0$ in (a) ,0.3 in (b), and 0.9 in (c),
respectively. The solid [cross (x)] line in nonvalence contribution
represents the full result of using true [average] $G_\pi$ value and the
dotted line represents the instantaneous part of the nonvalence contribution.
\label{SQD02}}
\end{figure}

\begin{figure}[t]
\centerline{\psfig{figure=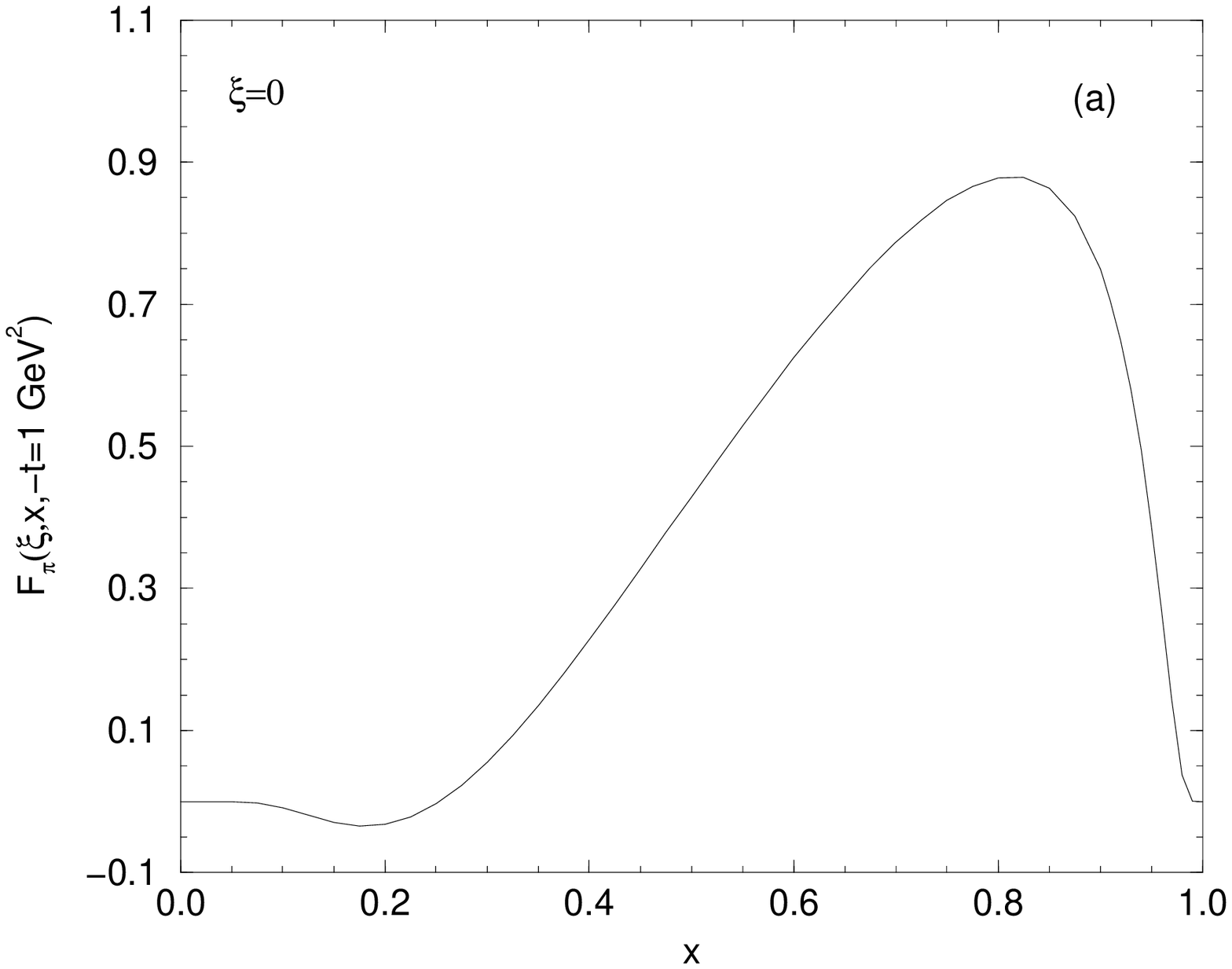,height=6.5cm,width=9cm}}
\centerline{\psfig{figure=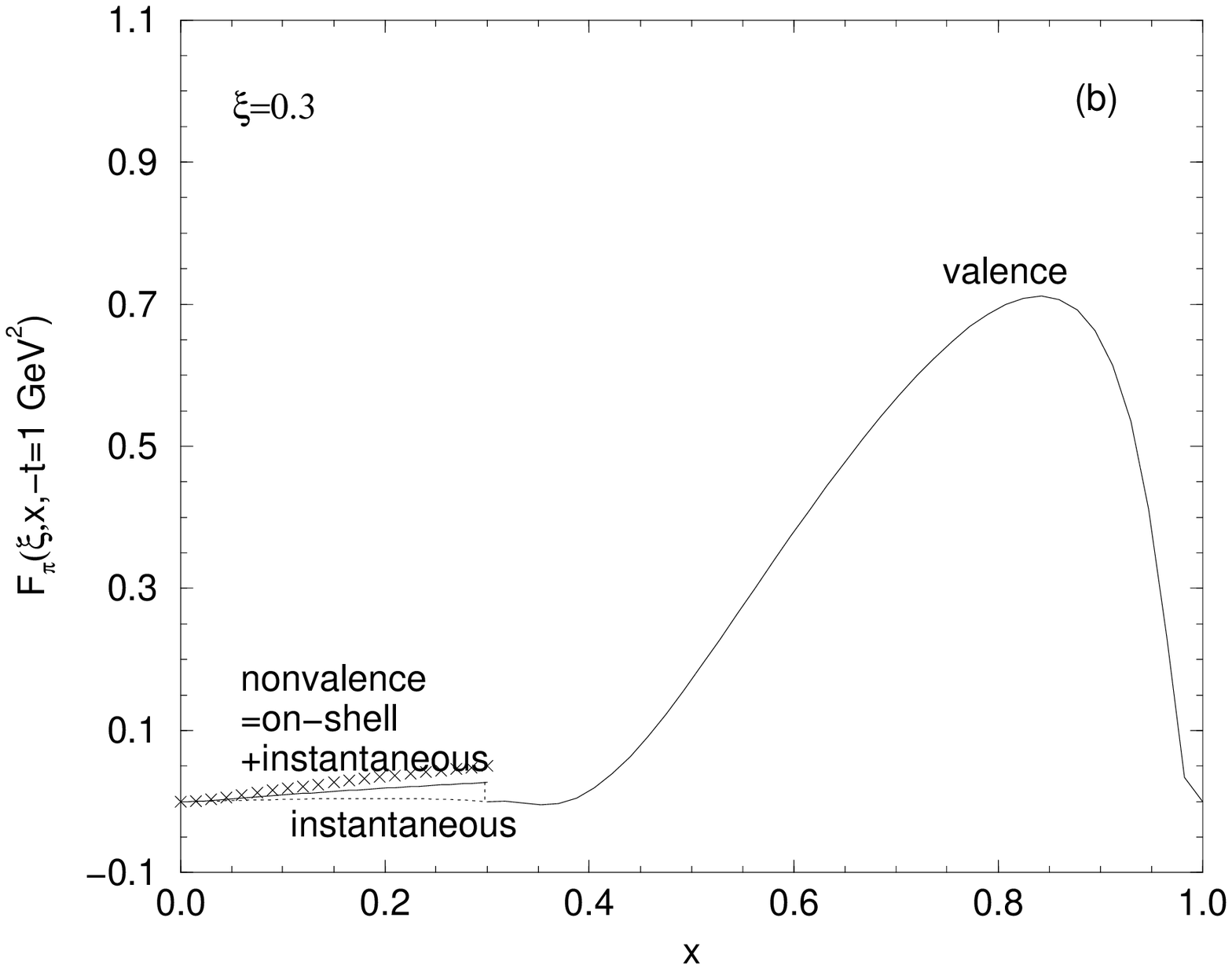,height=6.5cm,width=9cm}}
\centerline{\psfig{figure=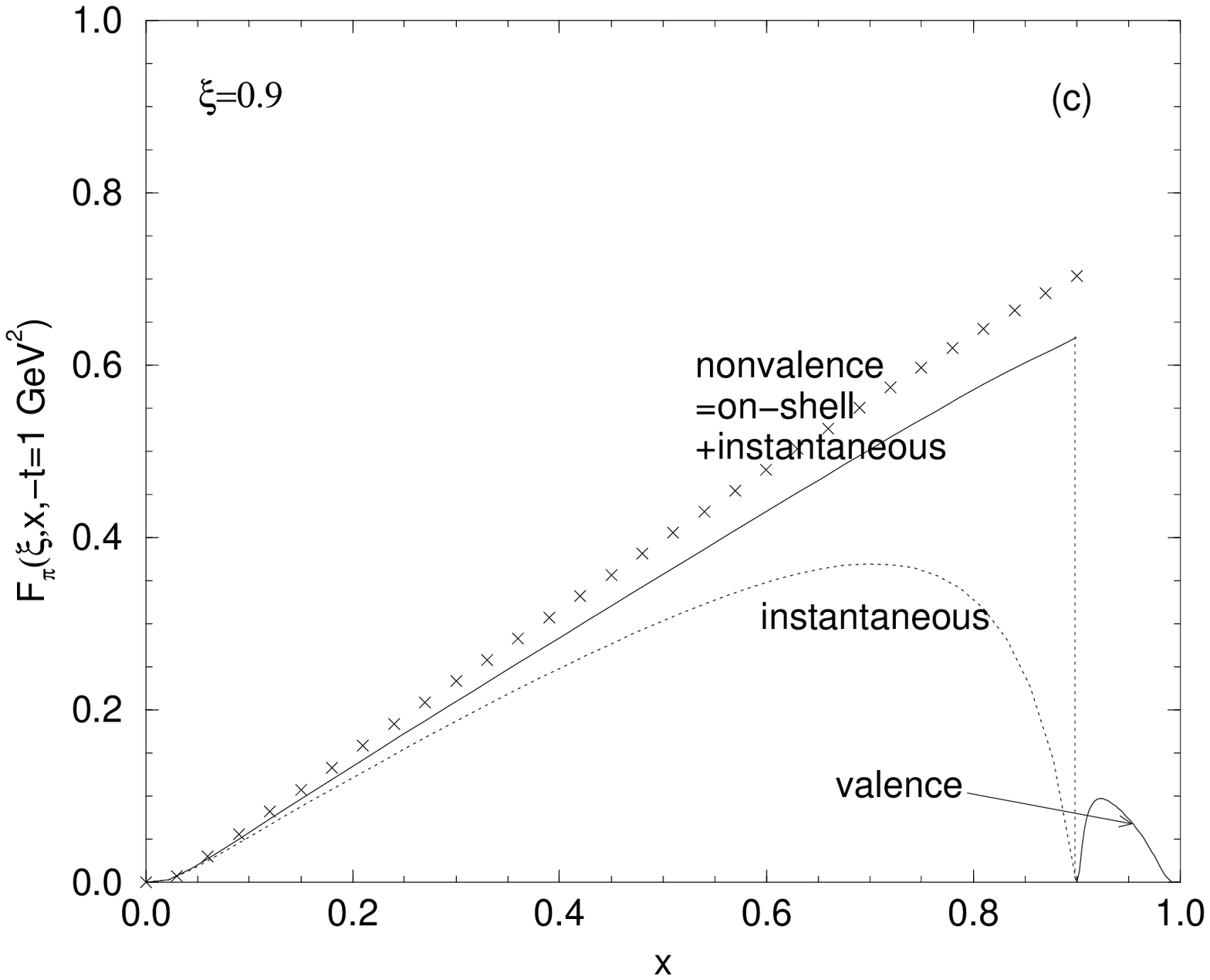,height=6.5cm,width=9cm}}
%\vspace{0.5cm}
\caption{Skewed quark distributions of the pion at
$-t=1$ GeV$^2$ with
different $\xi=0$ in (a), 0.3 in (b), and 0.9 in (c), respectively.
The same line code is used as in Fig.~\protect\ref{SQD02}.
\label{SQD1}}
\end{figure}

\begin{figure}[t]
\centerline{\psfig{figure=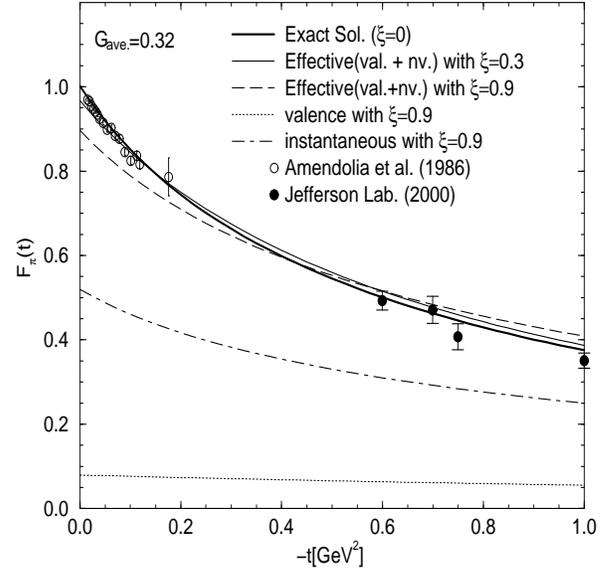,height=9cm,width=9cm}}
%\vspace{0.5cm}
\caption{The effective solution of pion form factor
using a common average $G_\pi=G_{\rm ave.}=0.32$ value
for $\xi=0.3$ (thin solid line) 0.9 (long-dashed line)
compared with the exact solution (thick solid line) as well as experimental
data~\protect\cite{Amen,JLab}. The dotted and dot-dashed lines represent
the valence and instantaneous contributions to the form factor for $\xi=0.9$
case, respectively.  \label{Piave}}
\end{figure}

\begin{figure}
\centerline{\psfig{figure=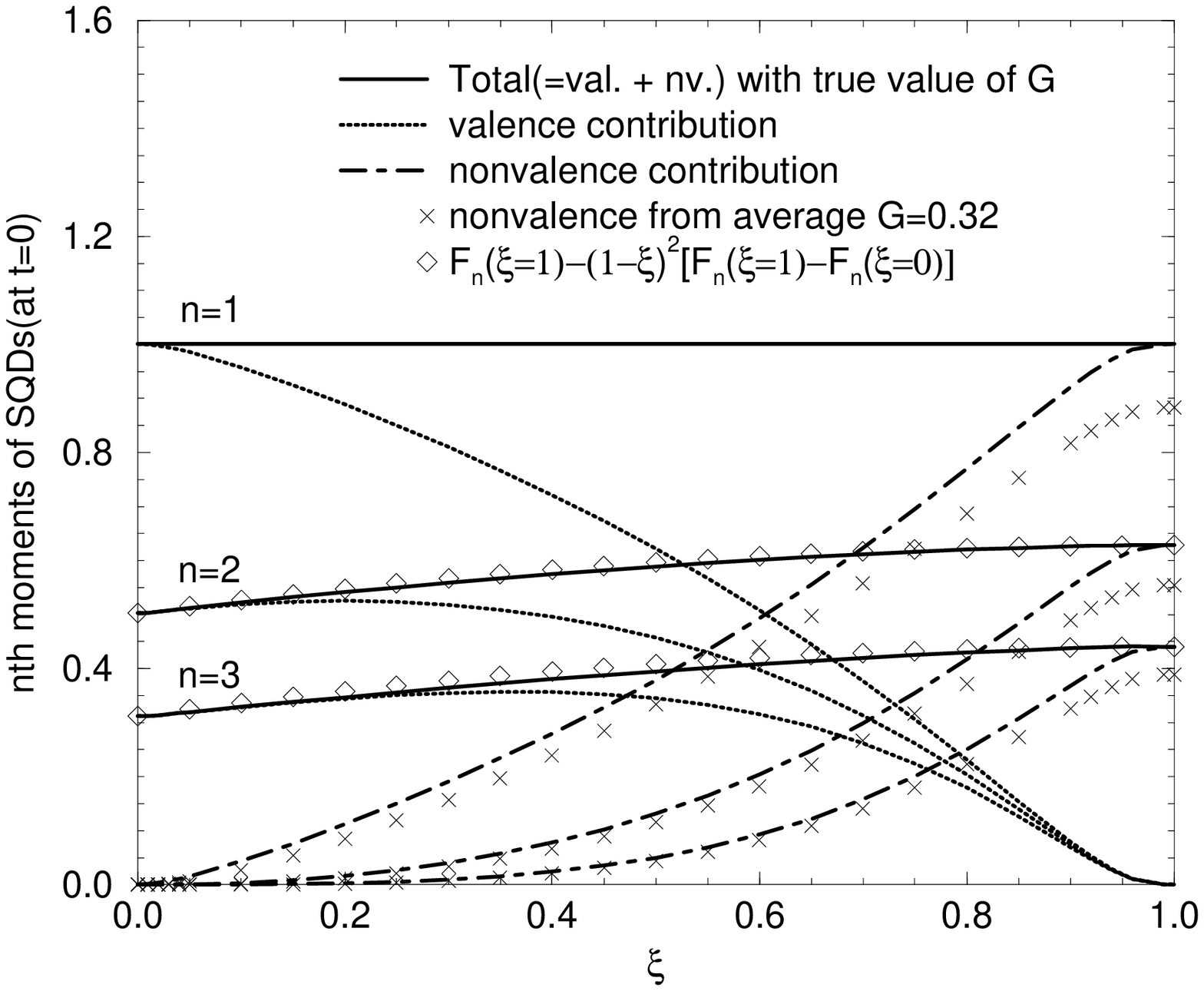,height=9cm,width=9cm}}
\caption{The $n$th moments $F_n(\xi,-t)$ of SQDs of the pion at 
$t=0$ using the true value of G. The total results(thick solid lines)
for $n=2$ and 3 are well fitted by the simple polynomial(diamond) 
in $\xi$.  For comparison, we include 
the nonvalence contributions obtained from the average 
$G_\pi=0.32$(cross lines).\label{Mellin}}
\end{figure}

\begin{figure}
\centerline{\psfig{figure=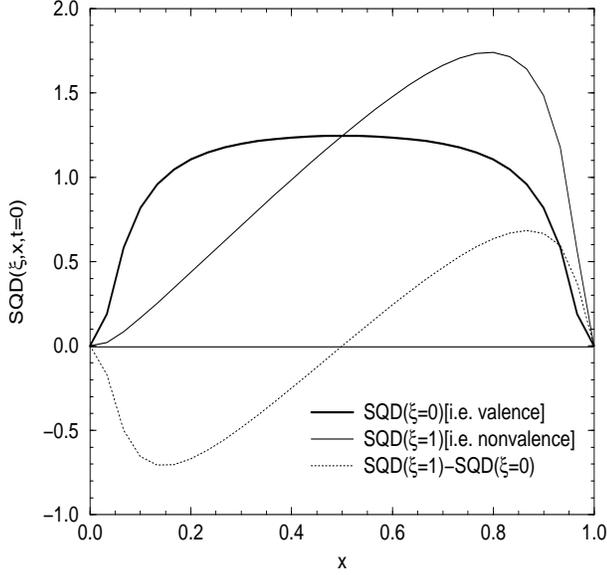,height=9cm,width=9cm}}
%\vspace{0.5cm}
\caption{The SQDs of the pion at
$t=0$ using the true value of G(=0.384). The thick solid, solid, and
dotted  lines represent ${\cal F}_\pi(\xi=0,x,t=0)$(valence),
${\cal F}_\pi(\xi=1,x,t=0)$(nonvalence), and their
difference ${\cal F}_\pi(\xi=1,x,t=0)-{\cal F}_\pi(\xi=0,x,t=0)$,
respectively.  \label{isoscalar}}
\end{figure}

The main approximation in our effective calculation is the
treatment of relevant operator ${\cal K}(x,{\bf k}_\perp;y,{\bf 
l}_\perp)$ in Eq.~(\ref{jnv}) connecting the one-body to three-body sector 
(see Fig.~\ref{Embed}) by taking a constant  
$G_\pi$ via $G_\pi\equiv\int[dy][d^2{\bf l}_\perp]
{\cal K}(x,{\bf k}_\perp;y,{\bf l}_\perp)\chi_{(2\to 2)}(y,{\bf l}_\perp)$,
which in general depends on $x$ and ${\bf k}_\perp$.
The reliability of this constant approximation was checked by examining
the frame-independence of our numerical results  
using the sum rule given by Eq.~(\ref{2sum}), i.e.
the exact results of ${\cal F}_\pi(\xi,x,t)$ and $F_\pi(t)$ 
obtained from the true values of
$G_\pi$ given by Fig.~\ref{Gfactor} were compared with those
obtained from our single averge value of $G_\pi=0.32$ for all
$(\xi, t)$. The numerical results of our constant $G_\pi$ prescription 
have shown definite improvement (better than 90 $\%$ accuracy for
$\xi\lsim 0.9$) to restore the
frame-independence of our model (see Figs.~\ref{SQD02}-~\ref{Piave})
and seemed to be a quite reliable approximation.
Our model also satisfies the polynomiality conditions for $n=2$ and $3$
moments of the SQDs (See Fig.~\ref{Mellin}) and our value of
$F_3^v$ is in good agreement with the QCD sum-rule result with 
non-local condensates~\cite{Bel}. Moreover, the ordinary quark 
distribution amplitude of the pion ($\phi_\pi(y=2x-1)$) can be obtained
from the two-pion distribution amplitude in $\xi=0$ and $t=0$ limit
where one of the produced pions becomes soft. Our result of $\phi_\pi$
is very close to the asymptotic quark distribution amplitude consistent
with the CLEO measurements~\cite{CLEO}. 
The isosinglet skewed quark distribution amplitude in the pion 
(See Fig.~\ref{isoscalar}) is also consistent with the result satisfying
the low-energy soft pion theorem~\cite{PW}.

For the model improvement, however, it would be necessary to consider
not only the dynamical quark mass discussed in our previous 
work\cite{KCJ} but also the B-S amplitude for the quark-gauge boson vertex 
to remove the apparent discontinuity at $x=\xi$ as shown in 
Figs.\ref{SQD02} and~\ref{SQD1}. 
Especially, in exploring the new physics associated with the SQDs for the 
treatment of the pion form factor, we note from Eq.~(\ref{SPD})  that the 
definition of ${\cal F}_{\pi}$ involves the matrix element of the 
light-front operator
present in the three-point approach to quark distributions of hadrons in
scaling regions. As was shown in Ref.~\cite{jk} the treatment of such
operators can lead one to consider nonlocal quark condensates, which can
introduce nonperturbative QCD structure in the quark-gauge boson vertex.
This will be a subject of future research by the authors.
\begin{center}
{\large\bf Acknowledgements}
\end{center}
The work of HMC and LSK was supported in part by the NSF
grant PHY-00070888 and that of CRJ by the US DOE under grant
No. DE-FG02-96ER40947. The North Carolina Supercomputing Center and 
the National Energy Research Scientific Computer Center are also
acknowledged for the grant of Cray time. 
%\begin{references}

\newpage

\mediumtext
\begin{table*}[t]
\caption{ The $n$th moments $F_n(\xi,x,t=0)$ of the SQDs for the pion
where $F^v_n=F_n(\xi=0)$ and $F^{nv}_n=F_n(\xi=1)$ represent
pure valence and pure nonvalence contributions, respectively.}
\begin{tabular}{ccccccc}
Model & $F^v_2[F^{nv}_2]$ & $F^v_3[F^{nv}_3]$ & $F^v_4[F^{nv}_4]$
& $F^v_5[F^{nv}_5]$ & $F^v_6[F^{nv}_6]$ & $F^v_7[F^{nv}_7]$\\
\hline
Ours& 0.503[0.623]  & 0.312[0.433] & 0.215[0.322]
& 0.16[0.25]    & 0.123[0.201] & 0.098[0.165] \\
\protect\cite{PW}  &    & 0.25[--] &   &   &   & \\
\protect\cite{Bel}  &    & 0.29[--] &   &   &   & \\
\end{tabular}
\end{table*}

\begin{table}
\caption{ The $n$th moments $\la y_n\ra$ of the ordinary
quark distribution amplitude $\phi_\pi (y)$ for the pion.}
\begin{tabular}{ccccccc}
Model & $\la y_1\ra$ & $\la y_2\ra$ & $\la y_3\ra$ & $\la y_4\ra$
& $\la y_5\ra$ & $\la y_6\ra$\\
\hline
Ours& 0.0  & 0.239 & 0.0 & 0.109 & 0.0 & 0.062 \\
\protect\cite{CZ} & 0.0 & 0.43 & 0.0 & 0.24 & 0.0 & 0.15 \\
\protect\cite{MR} & 0.0 & 0.25 & 0.0 & 0.12 & 0.0 & 0.07 \\
\protect\cite{JCC}$^c$ & 0.0 & 0.25 & 0.0 & 0.11 & 0.0 & 0.06 \\
\end{tabular}
$^a$ For the gaussian parameter $\beta=0.36$ GeV in Ref.~\cite{JCC}.
\end{table}
\end{document}